\documentclass{osa-article}

\journal{oe}

\usepackage{ulem}
\usepackage{multirow}
\usepackage{cite}

\begin{document}

\title{End-to-end topology for fiber comb based optical frequency transfer at the $10^{-21}$ level}

\author{Erik Benkler,\authormark{1,*} Burghard Lipphardt,\authormark{1} Thomas Puppe,\authormark{2} Rafa{\l} Wilk,\authormark{2} Felix Rohde,\authormark{2} and Uwe Sterr\authormark{1}}

\address{\authormark{1}Physikalisch-Technische Bundesanstalt, Bundesallee 100, 38116 Braunschweig, Germany\\
\authormark{2}TOPTICA Photonics AG, Lochhamer Schlag 19, 82166 Gr{\"a}felfing (Munich), Germany}

\email{\authormark{*}erik.benkler@ptb.de} 


\begin{abstract}
We introduce a simple and robust scheme for optical frequency transfer of an ultra-stable source light field via an optical frequency comb to a  field at a target optical frequency, where highest stability is required, e.g. for the interrogation of an optical clock. 
The scheme relies on a topology for end-to-end suppression of the influence of optical path-length fluctuations, which is attained by actively phase-stabilized delivery, combined with common-path propagation. This approach provides a robust stability improvement without the need for additional isolation against environmental disturbances such as temperature, pressure or humidity changes. We measure residual frequency transfer instabilities by comparing the frequency transfers carried out with two independent combs simultaneously. Residual fractional frequency instabilities between two systems of $8\times10^{-18}$ at 1~s and $3\times10^{-21}$ at $10^5$~s averaging time are observed. We discuss the individual noise contributions to the residual instability. The presented scheme is technically simple, robust against environmental parameter fluctuations, and enables an ultra-stable frequency transfer, e.g. to optical clock lasers or to lasers in gravitational wave detectors.
\end{abstract}

\section{Introduction}
Many precision metrology experiments rely on laser light with ultra-low optical frequency noise. A low frequency instability is particularly required at averaging times between 100~ms and 10~s, e.g. for atom-interferometer based gravitational wave detection~\cite{hog16, kol16, gra18a} or for quantum projection noise limited interrogation of optical lattice clocks \cite{alm15,oel19}, ion clocks \cite{san19}, or multi-ion clocks \cite{did19}.

The best cavity-stabilized lasers yield frequency instabilities as low as $4\times10^{-17}$ at averaging times between 100~ms and several ten seconds \cite{mat17a}, and further improvements are expected in the future from a silicon (Si) cavity with crystalline mirror coatings at low temperature of 4~K~\cite{oel19}, or from alternatives, such as QED systems \cite{chr15,nor18} or spectral hole-burning in cryogenic crystals \cite{coo15}. 

Often, the optimum performance of a reference system can only be implemented at a source wavelength (e.g. 1.5 $\mu$m for a Si cavity) distinct from the target wavelength, at which the ultra-stable signal is required (e.g. at an optical clock transition in the visible range). Thus usually the frequency stability from the light field at the source wavelength is transferred to the target wavelength using an optical frequency comb \cite{gro08, yam12, hag13}. 

The optical frequency comb provides a phase-coherent link between the fields in the different spectral regions. The beat notes between the comb modes and the cw fields carry the information about the relative frequency fluctuations between the source and target cw light field. These signals can thus be employed for phase-locking the target field such that ideally all relative fluctuations are eliminated and the stability of the source is transferred to the target field with high fidelity. 
In the ideal case, no additional technical noise should be added during the frequency transfer via the comb, besides quantum noise of the modelocked laser generating the frequency comb \cite{pas05} and quantum noise entering in the supercontinuum generation \cite{cor03, hav04a,dud06}. In reality, however, such excess noise occurs, with the largest contribution originating from optical path length fluctuations during the delivery of the cw and comb fields up to the point where they are superimposed for the generation of their beat notes. Such path segments, which contribute to the transfer excess noise due to differential optical length variations between the paths of the fields at the source and target wavelengths, will be referred to as \textit{uncompensated paths} in this paper. Often, the fields are delivered via optical fibers. The influence of various environmental parameter fluctuations on frequency transfer via fibers has been investigated in~\cite{wad19}.

The goal of this paper is to report a simple-to-implement, universal and robust method for the end-to-end elimination of uncompensated paths in a frequency transfer setup. The method relies on three essential elements:

(i) The cw fields are delivered from the source and target lasers over separate paths using well-established active frequency and phase stabilization techniques~\cite{ma94,wil08a}. The crucial feature of our approach is that the two phase-stabilized delivery systems share a common reference plane at their destination. This requires that the source and target fields are superimposed with each other before they reach the common semitransparent reference mirror defining this destination plane.

(ii) From the common reference plane onwards, the cw fields at the source and target wavelength propagate on a common path, up to the beam combiner, where they are superimposed with the comb fields. 

(iii) The spectral parts of the comb at the source and target wavelengths propagate on a common path as well before they are superimposed with the cw fields. Essentially, this means that they stem from a single branch of the comb generation system. Usually, in Er:fiber based frequency combs, light at the fundamental seed wavelength of $(1560 \pm 20)$~nm is emitted from every branch, along with the comb field at the target wavelength. This means that their comb lines have fixed phase relations to each other even if the comb spectral envelope vanishes between the two spectral regions. We will later discuss how our concept can be applied to source wavelengths outside the seed comb spectrum near 1560~nm.

Preferably, the common-path propagation of the fields at the two wavelengths is implemented in free space optics (air or even vacuum), but not in fibers, where the dispersion would have a larger effect.

During the past few years, several approaches for a stable frequency transfer via optical frequency combs were investigated. They can roughly be classified into two alternatives, often called the single- and the multi-branch approach. 
The single branch approach uses a broadband comb spanning over all required source and target wavelengths. The broadband comb either originates from a single branch of the comb generation system \cite{nic14,joh15,yao16,leo17,ohm17} or is generated in two distinct branches which are actively stabilized to each other \cite{rol18}. However, broadband supercontinuum generation from a narrow seed comb often involves multiple nonlinear conversion mechanisms with different spatio-temporal sensitivities. Thus the relative phases between the interfering different mechanisms are hard to control, which can lead to large fluctuations. As a result, robust and stable long-term operation of such broadband combs is hard to achieve. Furthermore, it is challenging to optimize the mode power in several spectral regions simultaneously.

The multi-branch approach instead uses separate branches for the generation of comb lines in the different spectral regions required. This allows individual engineering and optimization of the nonlinear processes at each target wavelength and thus higher amplitude stability of the comb lines. However, this comes at the expense that the path lengths of the independent branches can experience uncorrelated fluctuations, e.g. induced by variations of environmental parameters. Hitherto, this has mainly been suppressed engineering-wise by isolation against the influence of environmental parameters, under typical laboratory conditions down to instability levels around $10^{-16}$ at 1~s average time and $10^{-18}$ at averaging times longer than a few hours.

Recently, a combination of the multi- and single-branch approach has been used for stable frequency transfer~\cite{kas18}. Here, as in our approach, each branch is referenced to a common 1542~nm cw light field, and the seed comb field around 1542~nm propagates on a common path with the field at the target wavelength in each branch. In this sense, for each branch, a "single-branch" frequency transfer between 1542~nm and the target wavelength is carried out. This allows to gain the necessary information about the relative path length fluctuations between the multiple branches. 

In~\cite{kas18}, this information is used to actively stabilize all branches mutually via fiber stretchers as actuators. This active stabilization of separate branches enables a stable frequency transfer between optical frequencies, which are both outside the spectrum of the seed comb around 1560~nm. However, the actuators have a limited control bandwidth. In our approach, we eliminate the relative fluctuations by processing of the beat signals in the RF domain instead of actively stabilizing the path lengths. This requires no fiber stretcher actuators and allows a much higher control bandwidth for the elimination of path length fluctuation induced transfer instabilities between separate branches.

To the best of our knowledge, all hitherto reported frequency transfer setups involve some residual uncompensated paths. These uncompensated paths mostly result from the fact that the cw fields are delivered via actively phase-stabilized paths, but end at separate reference planes. Some approaches implement additional measures like evacuation~\cite{nic14} or dichroic heterodyne detection~\cite{giu19} to suppress the influence of path length fluctuations on these residual uncompensated paths. Our simple approach by principle completely avoids  uncompensated paths without the need for additional sophisticated technical requirements.

\section{Setup for end-to-end suppression of path-length induced frequency transfer instabilities and for frequency transfer performance measurement}
\label{sec:RPLC}

The frequency comb provides a phase-coherent link between two light fields at separate wavelengths, with the phases of the $m$-th comb line given by
\begin{equation}
    \varphi_m(t)=\varphi_\mathrm{CEO}(t)+m\varphi_\mathrm{rep}(t).
\end{equation}
Fluctuations of the comb's carrier-envelope offset (CEO) frequency $\nu_\mathrm{CEO}$ and repetition rate $f_\mathrm{rep}$ must be eliminated, because they would deteriorate the frequency transfer.
Usually, two methods are employed which can even be combined with each other. As first option, $\nu_\mathrm{CEO}$ is tightly locked to an RF reference signal, and $f_\mathrm{rep}$ is actively stabilized by a tight phase lock between the beat signal of the stable source cw field with the comb and an RF reference. This requires high-bandwidth CEO and repetition rate actuators.
The second option is to eliminate $\nu_\mathrm{CEO}$ and $f_\mathrm{rep}$ in the RF-domain using the transfer oscillator technique, which provides a ``virtual beat'' between the two cw fields~\cite{tel02b}.
Both options require a measurement of the beat notes between the cw fields and the comb, and of $\nu_\mathrm{CEO}$ using the $f-2f$ or similar technique~\cite{tel99}.

Alternatively, instead of locking the CEO frequency, it can be eliminated directly in the optical domain by Difference Frequency Generation (DFG) between two spectral sections of an initial comb with nonzero CEO frequency, which are an octave or more apart from each other, resulting in a CEO-free comb~\cite{kra11, pup16}. In contrast to the traditional approach of phase-locking the CEO beat signal, the DFG process eliminates CEO frequency fluctuations at Fourier frequencies up to the repetition rate~\cite{lie17}. Furthermore, it is more robust, does not require lock acquisition and does not produce servo bumps. Re-amplification with an EDFA after the DFG stage yields equal power levels as with the $f-2f$ approach. In principle, the additional nonlinear processes and the reamplification can introduce additional fundamental and technical noise. However, the integrated carrier-envelope phase noise of the DUT used in our experiments is comparable to the best reported values~\cite{tre19}.

In this work we use the transfer oscillator technique applied to a CEO-free DFG-comb (TOPTICA DFC CORE+) as the ``device under test'' (DUT). The beat note between the DUT comb and the cw source laser stabilized to the Si cavity is used to phase-lock the repetition rate of the DUT comb. We also performed the same measurements when the DUT comb repetition rate was RF-locked to an active hydrogen maser causing a less stable repetition rate. Nevertheless, similar frequency transfer instabilities as those we report here for the optically locked DUT comb were observed in the RF-locked case. To assess the performance of the frequency stability transfer in an out-of-loop measurement, we use a second, independent comb with its CEO frequency self-referenced using the $f-2f$ technique (Menlo FC-1500-ULN), combined with the transfer oscillator technique as well to simultaneously perform a ``reference'' frequency transfer. To determine the excess noise entering during the frequency transfer, we compute the difference between the fractional optical frequency ratios
\begin{equation}
    y_\mathrm{DUT-ref}(t)=\frac{ \left.\frac{\nu_\mathrm{target}}{\nu_\mathrm{source}}\right|_\mathrm{DUT}(t)-\left.\frac{\nu_\mathrm{target}}{\nu_\mathrm{source}}\right|_\mathrm{ref}(t)}{\left.\frac{\nu_\mathrm{target}}{\nu_\mathrm{source}}\right|_\mathrm{nom}},
    \label{eq:ratiodiff}
\end{equation}
where $\nu_\mathrm{source}$ and $\nu_\mathrm{target}$ are the source and target absolute optical frequencies, and the index nom indicates the ratio between some nominal absolute optical frequency values, which are close to the real absolute frequencies. The normalization by the nominal ratio yields the fractional frequency ratio.
The other optical frequency ratios are computed from the beat frequencies counted at the DUT and reference system, respectively, using the transfer oscillator scheme. Thus we compare the optical frequency ratios measured simultaneously with the two frequency transfer systems. As a consequence, fluctuations of the cw optical frequencies of source and target lasers cancel out and only residual excess noise due to the two transfers is left. Finally, we compute from $y_\mathrm{DUT-ref}$ the modified Allan deviation $\mathrm{mod}\,\sigma_{y_\mathrm{DUT-ref}}$ as a measure for the instability. Please notice that the differentiation between DUT and reference system is only formal and that we can only measure their combined instability. A three-cornered hat measurement~\cite{ver16a} using three transfer systems could yield information about the performances of the individual frequency transfer systems, but is not within the scope of this paper.

\begin{figure}[htbp]
\centering
\includegraphics[width=\linewidth]{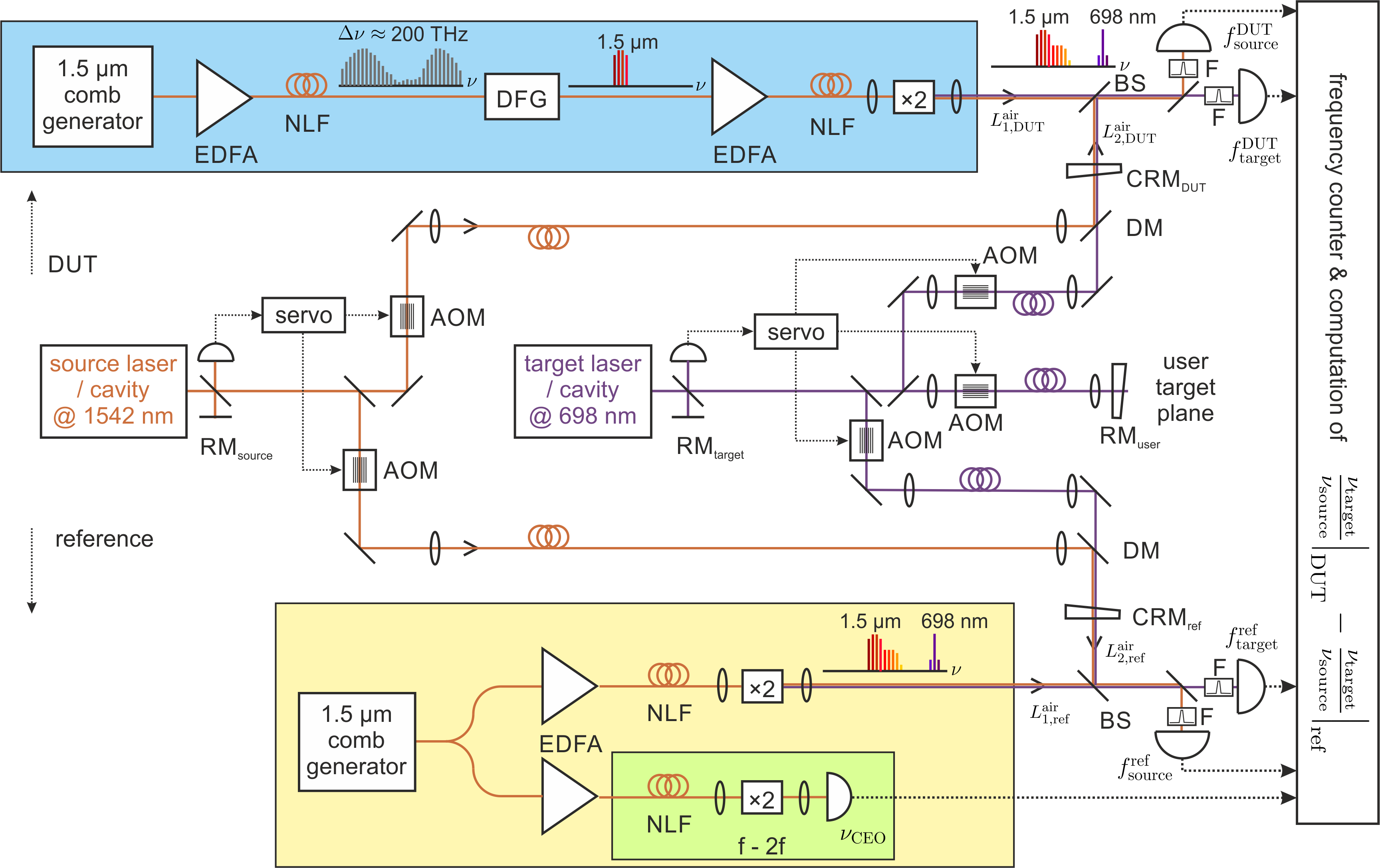}
\caption{Schematic experimental setup for the evaluation of the frequency transfer instability: CP configuration, free of uncompensated paths in both the DUT system (upper half) and reference system (lower half). The DUT comb generation system (blue box) generates a CEO-free optical frequency comb, while the reference comb generation system (yellow box) provides a CEO frequency measurement based on a common-path $f-2f$ interferometer (green box). NLF: Highly nonlinear fiber, EDFA: Erbium-doped fiber amplifier. See text for further details of the comb generation systems.
The source cw field (brown lines) is emitted by a laser stabilized to an ultra-stable Si cavity at 1542~nm, and the target cw field (purple lines) is emitted by a Sr lattice clock laser pre-stabilized to a 48~cm long cavity at 698~nm~\cite{hae15a}. By active path length stabilization, the phases at the planes defined by the near-end reference mirrors $\mathrm{RM_{source}}$ and $\mathrm{RM_{target}}$ have a fixed relation to the phases at the far-end planes $\mathrm{CRM_{ref/DUT}}$ and $\mathrm{RM_{user}}$, respectively. The user target plane $\mathrm{RM_{user}}$ could e.g. be located close to the atoms of a Sr lattice clock to transfer the stability of the source laser to the location of the clock interrogation. RF signal paths are indicated by dotted lines. The beat signals between cw fields and combs are detected by photodiodes at RF frequencies $f_\mathrm{source}$, $f_\mathrm{target}$ and $\nu_\mathrm{CEO}$ (ref system only), pre-filtered using PLL tracking filters, and counted with a dead-time-free frequency counter in $\Lambda$-mode~\cite{kra04a}, such that the data can be evaluated in a post-processing step.}
\label{fig:cp_setup}
\end{figure}

\begin{figure}[htbp]
\centering
\includegraphics[width=\linewidth]{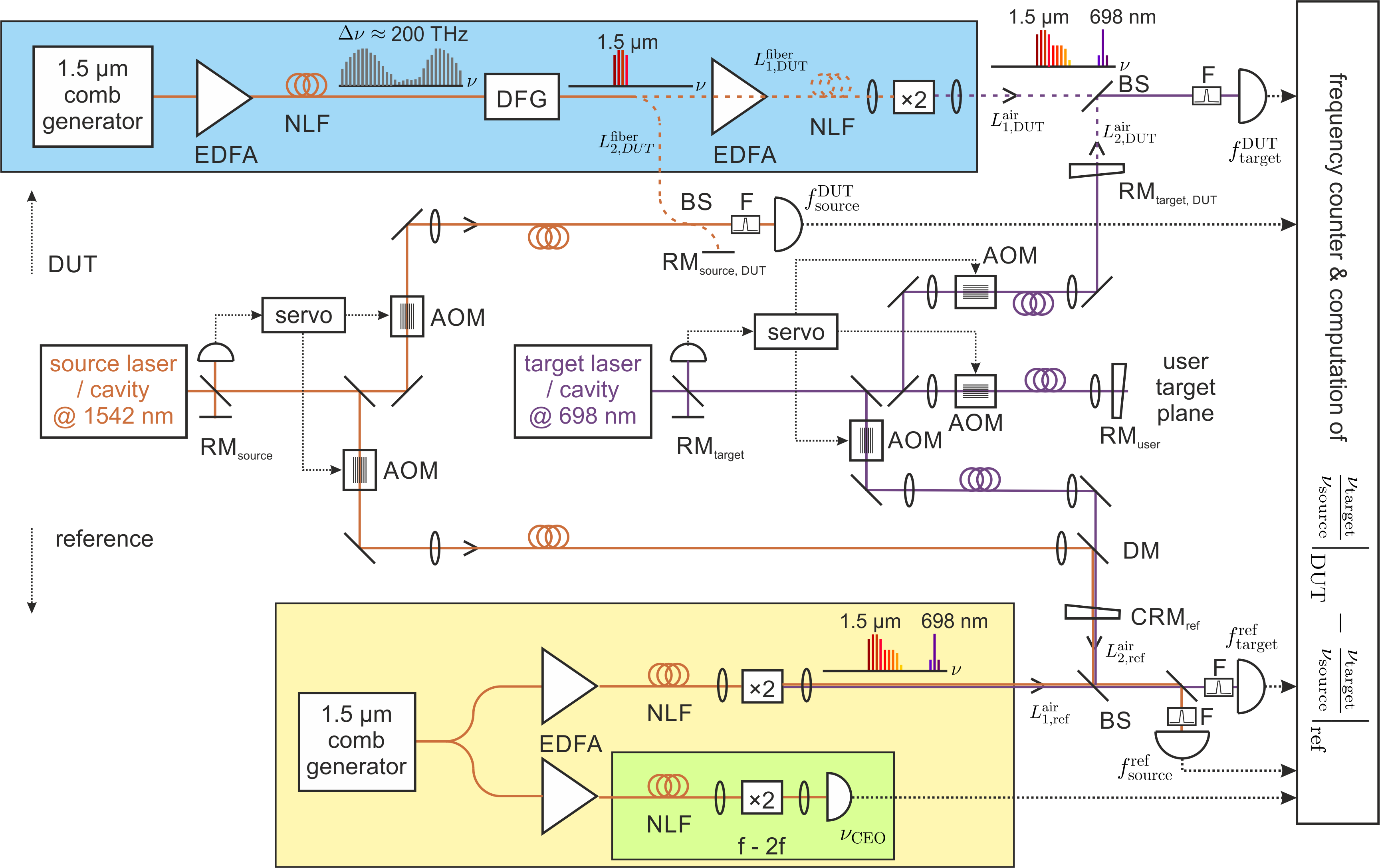}
\caption{Schematic experimental setup for the evaluation of the frequency transfer instability: NCP configuration, containing uncompensated path segments in the DUT system, as indicated by dashed lines. Compensated optical paths are shown as solid lines. In contrast to the CP case, the superposition of the source field with the DUT comb is implemented completely as fiber optics, and using a separate branch. As in the CP case, the reference system does not comprise uncompensated paths.}
\label{fig:ncp_setup}
\end{figure}

To demonstrate the robust suppression of optical path-length-induced noise using our end-to-end approach, we measure the transfer instability in two configurations, shown in Fig.~\ref{fig:cp_setup} and~\ref{fig:ncp_setup}. 

The two configurations only differ in the way the beat signals are generated in the DUT system: We abbreviate the first configuration as ``CP'' (compensated paths), because the uncompensated paths are completely eliminated in the DUT system as well as in the reference system, as shown in Fig.~\ref{fig:cp_setup}. The following description is identical for the DUT (upper half) and reference (lower half) system if not mentioned otherwise.

For the end-to-end suppression of path-length-induced frequency instabilities, we overlap the source and target cw fields on the dichroic mirror (DM), before they reach the common reference mirror (CRM). The cw light fields reflected by the CRM propagate back to their sources, where their phase is interferometrically detected and actively stabilized relative to the respective near-end reference mirror using acousto-optical modulators (AOMs) as actuators~\cite{wil08a}. Using different AOM shift frequencies, the RF beat signals for the active stabilization of the individual paths can be separated by filtering. Within the servo bandwidth of the active phase stabilization loops, the source and target phase at the plane defined by the CRM thus have a fixed relation to the phase at the reference mirror near the respective cw laser source. Hence at the CRM, the length fluctuations between the two cw paths are completely compensated. Subsequently, the fields transmitted through the semi-transparent CRM propagate on a common path up to the beam splitter (BS), where they are superimposed with the comb field. We use a comb field from a single branch of the comb generation system, which is possible even with branches, which are narrow-band optimized for a certain target wavelength, because the residual $1560$~nm seed comb around the source wavelength is usually emitted along with the comb at the target wavelength. The frequency doubling stages for generation of the comb fields at the target wavelength (698~nm in our experiments) are implemented by free space PPLNs in both the DUT and reference system. After their superposition with the cw fields on the beam splitter BS, the two spectral regions are safely separated into two detection paths, e.g. by a dichroic beam splitter. As usual, a narrowband optical filter F rejects unwanted spectral components besides the cw line and a few hundred comb lines around it. 

By this combination of actively stabilized paths and common path propagation, critical paths are compensated completely from the near ends to the far ends. In our experiments, only $L_\mathrm{1,DUT}^\mathrm{air}=35$~cm, $L_\mathrm{2,DUT}^\mathrm{air}=30$~cm free space paths in air in the DUT system, and $L_\mathrm{1,ref}^\mathrm{air}=35$~cm, $L_\mathrm{2,ref}^\mathrm{air}=$ 110~cm (see Figure~\ref{fig:cp_setup} for their locations) in the ``reference'' system are left, where the fields propagate on a common path, but at different wavelengths. On such sections, path length fluctuations are indeed suppressed, but only partially and limited by the dispersion of the air.

The second configuration, termed ``NCP'' (non-compensated paths), in contrast to the CP case uses two separate branches of the DUT comb generation system for the generation of the beat notes with the cw fields. Hence it contains $L_\mathrm{1,DUT}^\mathrm{fiber}=12.53$~m, $L_\mathrm{2,DUT}^\mathrm{fiber}=6.53$~m uncompensated polarization maintaining fiber path and $L_\mathrm{1,DUT}^\mathrm{air}=35$~cm, $L_\mathrm{2,DUT}^\mathrm{air}=30$~cm uncompensated free space path in air, where the effect of length fluctuations is not suppressed due to common mode rejection. The uncompensated paths are indicated by dashed lines in Fig.~\ref{fig:ncp_setup}. The reference system is identical as in the CP case.

The DUT system, the reference system, the source laser and the target laser are located in separate rooms, such that common mode effects are excluded to a large extent.

For the reference system an additional frequency $(\nu\mathrm{CEO})$ needs to be counted, in contrast to the DFG-based DUT comb system, where only 2 frequencies need to be counted. The $f-2f$ based detection of the CEO frequency is performed in a separate path of the reference comb generation system. Optical path length fluctuations, however, are efficiently suppressed, first because a common-path $f-2f$ interferometer is used, and second because the CEO phase only depends on fluctuations of the dispersion along the paths, which are negligible compared to the direct impact of optical path length variations. 

The CEO beat signal is detected with a white phase noise floor of $S_\varphi\approx-90\,\mathrm{dBrad^2/Hz}$ (SNR of 43~dB in 100~kHz RBW). The RF beat signals at $f_\mathrm{source}$ and $f_\mathrm{target}$ are detected by photodiodes with a white phase noise floor better or equal to $S_\varphi\approx-80\,\mathrm{dBrad^2/Hz}$ (33~dB in 100~kHz) at both combs. These beat signals and the CEO beat signal are amplified and filtered using PLL tracking filters with a bandwidth of $f_\mathrm{track}\approx1$~MHz. These signals could in principle be processed by RF hardware following the approach from~\cite{tel02b} to generate a virtual beat signal at the frequency $f_\mathrm{transfer}= \nu_\mathrm{source}-\nu_\mathrm{target} \times m_\mathrm{source}/m_\mathrm{target}$ in real-time, where $m_\mathrm{source}$ and $m_\mathrm{target}$ are the comb line order numbers of the corresponding beat notes with the comb. The transfer beat signal can then be used to generate an error signal for phase-locking the frequency fluctuations of the target field at 698~nm to those of the source field at 1542~nm. We have shown in an independent measurement that the in-loop instability we achieve with this phase lock is sufficiently smaller than the observed transfer instabilities observed due to the other parts of the transfer setup. Furthermore, it is not necessary to actually carry out the phase lock for the CP-NCP comparison. For this reason, we substitute the phase lock by a post-processing analysis of the beat frequencies involved. 
For this purpose, the RF frequencies are counted using synchronous dead-time free frequency counters (K\&K FXE)~\cite{kra04a} referenced to a hydrogen maser. The counters with a fundamental gate time of $\tau_0=1$~ms are operated in $\Lambda$ frequency averaging mode with an averaging time of 1~s~\cite{rub05, ben15}. We post-process the recorded frequency time series on a 1~s grid to get the time-resolved frequency ratios between the source and target frequency measured at each of the two frequency combs, which we then plug into Eq.~(\ref{eq:ratiodiff}) and finally compute the modified Allan deviation as a measure of the combined excess instability.

The measured instability between both systems is shown in Fig.~\ref{fig:results} for the measurements in the NCP (blue line) and CP (orange line) configuration.
The instability measured in the NCP configuration starts at a relative frequency instability of $10^{-16}$~@~1~s. It increases to a maximum near $\tau\approx$~10~s. At longer averaging times, the NCP instability drops below $10^{-18}$ at $\tau>5000$~s. 

\section{Experimental results and discussion}
\label{sec:results}
\begin{figure}[htbp]
\centering
\includegraphics[width=\linewidth]{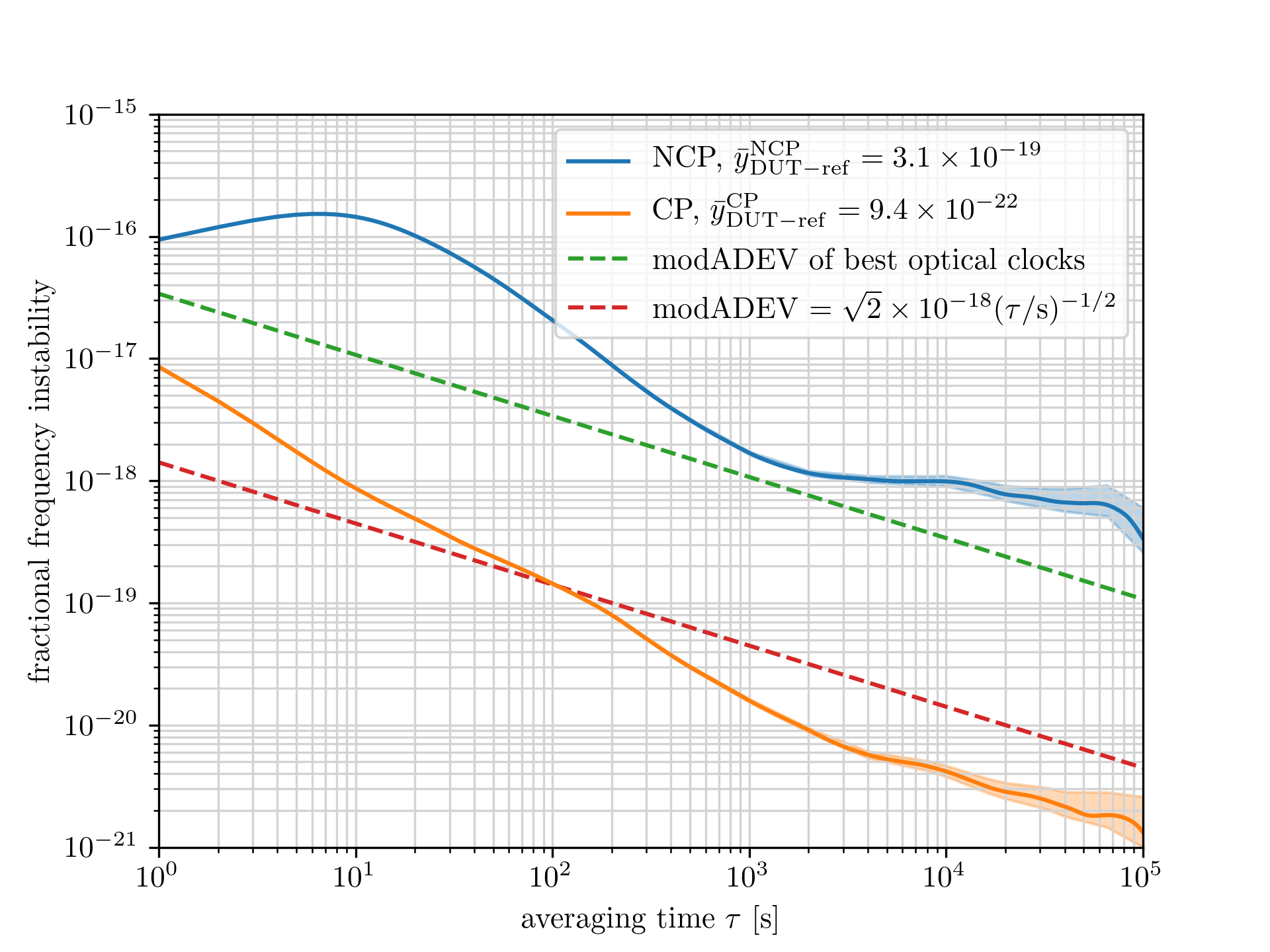}
\caption{Instability plot. Blue: Measured combined instability in the NCP configuration. Orange: Measured combined instability in the CP configuration. The shaded regions indicate $1\sigma$ confidence intervals estimated by the method of Greenhall and Riley~\cite{gre03b}. As a measure for the instability, the modified Allan deviation was chosen. For comparison, the green dashed line shows the modified Allan deviation $\mathrm{mod}\,\sigma_y(\tau)= 4.8\times10^{-17}/\sqrt{2\tau/\mathrm{s}}$ of the most stable optical clocks to date~\cite{oel19} and the red dashed line shows a modified Allan deviation of $\mathrm{mod}\,\sigma_y(\tau)=\sqrt{2}\times10^{-18}/\sqrt{\tau/\mathrm{s}}$. This corresponds to white frequency noise with an Allan deviation of $\sigma_y(\tau)=2\times10^{-18}/\sqrt{\tau/\mathrm{s}}$, which is a factor of 24 better than the best optical clocks. For measurement times $\tau > 100$~s, the statistical uncertainty of a frequency comparison between clocks with this performance would not be limited by the transfer via the comb. The $y_\mathrm{DUT-ref}$ time series data can be downloaded from PTB's open data repository~\cite{ben19a} for further analysis, such as estimation of the power spectral density or Allan deviation.}
\label{fig:results}
\end{figure}

In the CP case instead, the measured instability starts at $8\times 10^{-18}$ at $\tau=1$~s and continuously drops to below $3\times 10^{-21}$ at $10^5$~s including its uncertainty. By our end-to-end approach, the instability due to path length variations is suppressed by a factor of at least 200, but as the CP and NCP instabilities differ in their dependencies on the averaging time $\tau$, we conclude that the limiting processes are different in the two cases. The observed frequency transfer performance is as good as the best reported ones~\cite{nic14, leo17, rol18}, but requiring considerably less technical efforts, i.e., no vacuum, no active or passive temperature stabilization, and no broadband optical frequency comb. 

Besides the stability transfer performance, we can use the measurements to check the agreement between the mean values of the optical frequency ratios measured at the DUT and reference system, as given by the arithmetic mean of Eq.~(\ref{eq:ratiodiff}). We determine an average residual fractional frequency ratio difference of $\bar{y}\mathrm{_{DUT-ref}^{NCP}}= 3.1 \times 10^{-19}$ for the NCP case, and $\bar{y}\mathrm{_{DUT-ref}^{CP}}= 9.4 \times 10^{-22}$ for the CP case, which is compatible with zero within the $1\sigma$-uncertainties determined from the instability plots at the longest averaging time.
\begin{figure}[htbp]
\centering
\includegraphics[width=\linewidth]{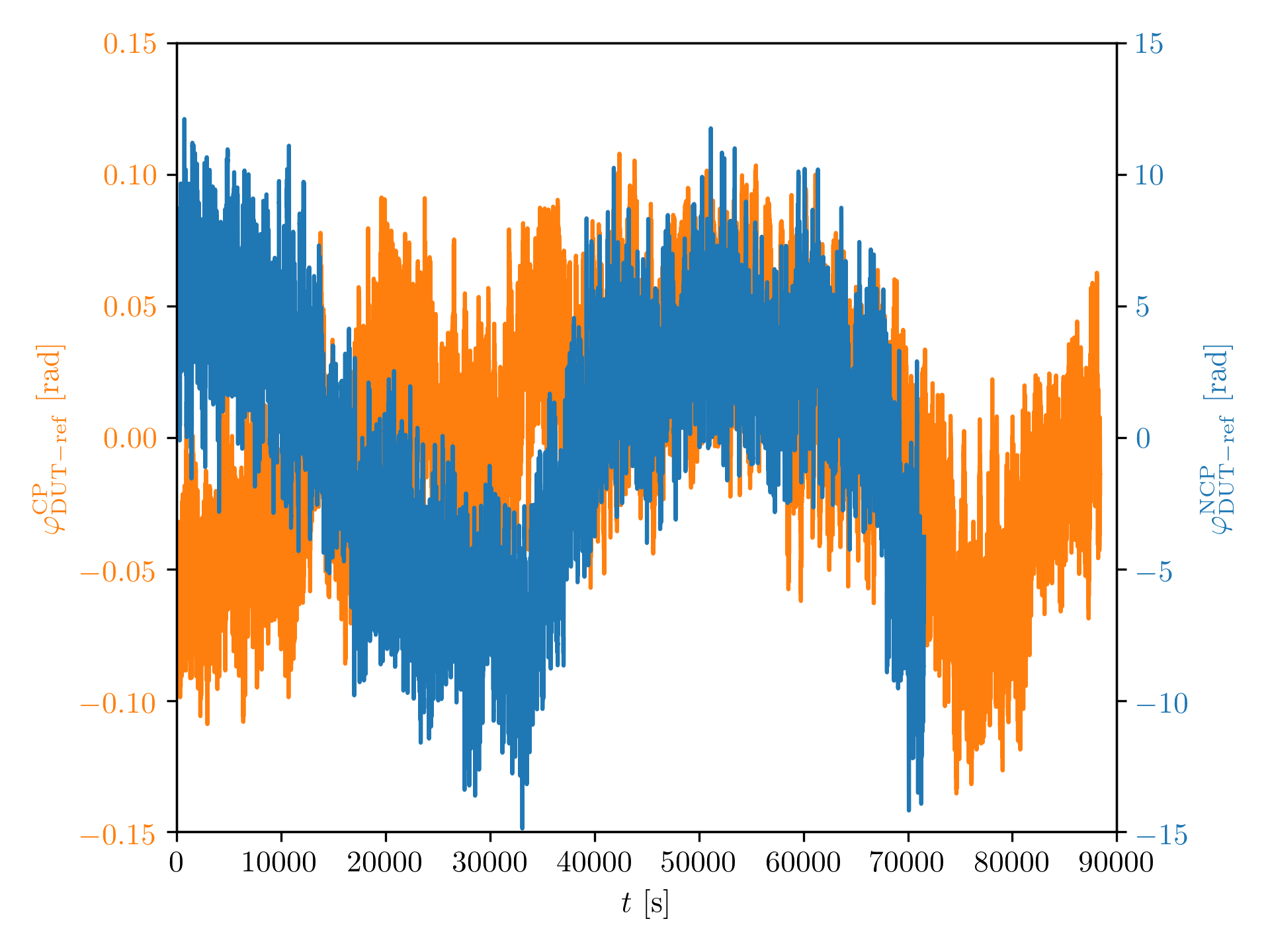}
\caption{Temporal phase evolution $\varphi_\mathrm{DUT-ref}(t)$ scaled to an optical carrier frequency of 194.4~THz during time intervals with continuous data. The CP (orange, left scale) and NCP (blue, right scale) measurements were not performed simultaneously.}
\label{fig:phase_evolution}
\end{figure}
It should be noticed that the $y_\mathrm{DUT-ref}$ data sets contain some gaps. These gaps have two origins of different nature: First, $y_\mathrm{DUT-ref}(t_i)$ data cannot be used at time stamps $t_i$ when a cycle slip occurred in one of the active path length stabilization PLLs or in the RF tracking filters. This typically leads to single missing values. The average rate of these short events is 35 cycle slips / day, corresponding to 0.04~\% of the data points. Second, the clock laser at the target wavelength sometimes fell out of the phase lock to its pre-stabilization cavity, which resulted in four longer gaps (average gap length of roughly 25000~s) only during the CP measurement. These events did not occur during the NCP measurement. Figure~\ref{fig:phase_evolution} shows the phase evolution $\varphi_\mathrm{DUT-ref}(t)$ during a continuous time interval without gaps (please take notice of the different scalings of the y-axis for the CP and NCP case).

In order to better understand the underlying processes contributing to the measured instability and frequency ratio offsets, we estimate typical fractional frequency instabilities resulting either from fundamental noise processes like shot noise in the detection of the beat notes on photodiodes, or from variation in environmental parameters such as temperature, relative humidity or pressure.

Fluctuations of optical path lengths lead to fluctuating Doppler frequency shifts and hence to fluctuations of the difference between the source and target frequency. 

The optical path length is given by the geometric path length $L$ times the effective index of refraction $n$ averaged along the path. A temporal variation of the optical path length causes a fractional Doppler shift 
\begin{equation}
y=\frac{\Delta \nu}{\nu_0}= \frac{1}{c} \left(L\frac{\mathrm{d}\,n}{\mathrm{d}\, t} + n\frac{\mathrm{d}\,L}{\mathrm{d}\, t}\right),
\label{eq:Dopplershift}
\end{equation}
where $c$ is the speed of light, $\nu_0$ is the optical carrier frequency and $\Delta\nu$ is the absolute frequency Doppler shift. 
Eq.~(\ref{eq:Dopplershift}) shows that such a Doppler shift can result from a variation of the refractive index, e.g. due to temperature ($T$) or pressure ($p$) fluctuations, or from a variation of the geometrical length, e.g. due to thermally or humidity-induced ($H$) expansion of optical fibers. We can thus derive the quantitative noise contributions from various individual environmental parameter data, on different path sections in the optical setup, e.g. those implemented as free space or in fiber optics, or on common or non-common paths at the same or at different wavelengths. The individual contributions on an optical path length $nL$ can be derived from the environmental parameters $\zeta = T,\, p,\, H,\, \dots$ by
\begin{equation}
y_{\zeta}= \frac{1}{c}
\left(
L\frac{\partial\,n}{\partial\,\zeta} 
+n \frac{\partial\,L}{\partial\,\zeta} 
\right)\frac{\mathrm{d}\,\zeta}{\mathrm{d}\,t}.
\label{eq:total_diff}
\end{equation}
Typical coefficients of single-mode fibers from literature are shown in Tab.~\ref{table1}. There is a large variation of these parameters depending on the details of the fibre type, coating and jacket. As a worst case estimate we have used the humidity coefficient of a humidity sensing fiber, which is significantly larger than the coefficient for standard single-mode fibers. Furthermore, our model uses these steady-state-coefficients and therefore does not take into account dynamics like the slow process of water diffusion into the fiber's polyimide coating (typical time constants on the order of an hour~\cite{gia01}, which becomes even longer if the fiber has a polymer jacket as in our experiments), which causes swelling of the polyimide layer and thus expansion of the fiber length. As a result, the humidity contributions are most likely over-estimated, especially at averaging times $\tau$ shorter than the indiffusion time. Within the accuracy of our model, we can use the simpler parameters of SMF28 fiber which are available in the literature to estimate the contributions in the PM fibers actually used in our experiment. The effective phase refractive index in SMF28 fiber (relevant for our estimates and not to be confused with the effective group index given directly in the SMF28 data sheet) has been computed with a freeware fiber simulation software~\cite{rpf}. For the computations, we assume a step-index profile with 4.1~$\mu$m core radius, the Sellmeier coefficients of pure fused silica for the cladding material index, and a 0.36\% index jump between cladding and core due to GeO$_2$ doping.

\begin{table}[htbp]
\centering
\caption{\bf Fiber parameters used for the calculation of path length fluctuations.}
\begin{tabular}{ |c|c|c| }
\hline parameter & value & reference \\ \hline
$\frac{1}{n}\frac{\partial n}{\partial T}$ & $10^{-5}/\mathrm{K}$ & \cite{lag81}\\
$\frac{1}{n}\frac{\partial n}{\partial p}$ & $-8 \times10^{-11}/\mathrm{Pa}$ & \cite{bud79}\\
$\frac{1}{L}\frac{\partial L}{\partial T}$ & $6.4\times 10^{-7}/\mathrm{K}$  & \cite{kue09}\\
$\frac{1}{L}\frac{\partial L}{\partial H}$ & $4.36\times 10^{-6}/\%$  & \cite{gia01}\\
$n_\mathrm{eff}(1397~\mathrm{nm})$ & 1.4483 & \multirow{2}{*} {\cite{mal65, smf28,rpf}}\\
$n_\mathrm{eff}(1542~\mathrm{nm})$ & 1.4463 &\\
\hline
\end{tabular}
\label{table1}
\end{table}
In coaxial RF cables, the temperature coefficient for fluctuations of the electrical length $L^*$ is $\frac{1}{L*}\frac{\partial L^*}{\partial T} < 10^{-4}/\mathrm{K}$ and the pressure coefficient is $\frac{1}{L*}\frac{\partial L^*}{\partial p} < 10^{-4}/\mathrm{Pa}$~\cite{czu11, lut89}. In analogy to Eq.~(\ref{eq:Dopplershift}), we can compute a relative frequency instability $y_\mathrm{RF}$. For a comparison with the effects of optical path length fluctuations,  $y_\mathrm{RF}$ must be rescaled to optical frequencies instead of RF frequencies by the factor $f_\mathrm{RF}/\nu_0\approx10^{-7}$. Since the RF cable lengths are comparable to the fiber lengths in our setup, the effects of electrical length fluctuations can be neglected with respect to the optical length fluctuations.

For free space paths, we model the refractive index of air $n(\lambda,T,p,H)$ and its dependence on environmental parameters by the improved Edl\'en formula~\cite{edl66,bir94}.  

To estimate the noise contributions for a given experimental setting, common-path and non-common path sections must be treated separately, and their types (e.g. fiber or free space / air) must be taken into account. Finally, the contributions from the separate paths must be added with correct signs. For example, for a beat signal between two fields which propagated along two separate paths, their differential phase must be considered. Furthermore, we must make an assumption on the correlations between the phase fluctuations accumulated during the propagation along the two separate paths. Since we use global, spatially independent environmental parameters only, it is a reasonable assumption that the fluctuations on the two paths are fully correlated. According to Eq.~(\ref{eq:total_diff}), contributions from individual paths at a single wavelength are proportional to their geometrical lengths. Hence, the differential phase is proportional to the difference between the two lengths, $L_\mathrm{diff}= L_1 - L_2$ in the case of full correlations. The relevant geometrical lengths used for the estimations are taken from our experimental setup and listed in table~\ref{table2}.
\begin{table}[htbp]
\centering
\caption{\bf Geometrical path lengths used for the estimation of the various instability contributions. They are chosen as in the experiment and their locations are indicated in Figures~\ref{fig:ncp_setup} and ~\ref{fig:cp_setup}.}
\begin{tabular}{|c|c|c||c|c|}
\hline
type & system & $\lambda$ [nm] & $L_\mathrm{1}$ [m] &$L_\mathrm{2}$ [m] \\
\hline\hline
NCP, air & DUT & 698 & 0.35 & 0.3 \\
\hline
NCP, fiber & DUT & 1542 & 12.53 & 6.53 \\
\hline\hline
\multirow{2}{*}{CP, air} & DUT  & \multirow{2}{*}{698 \& 1542} & 0.35 & 0.3\\
 & ref & & 0.35 & 1.1 \\
 \hline
\multirow{2}{*}{CP, fiber} & DUT  & \multirow{2}{*}{1397 \& 1542} & 0.02 & --\\
 & ref & & 0.04 & --\\
\hline
\end{tabular}
\label{table2}
\end{table}
In reality, the fluctuations are neither fully correlated nor fully uncorrelated and the correlation may depend on the Fourier frequency of the fluctuations. Since temporal fluctuations on time scales longer than the relaxation time to spatial equilibrium of an environmental parameter have a more homogeneous spatial distribution, the fluctuations in two separate paths tend to be more correlated on long term than on short term. From these considerations should be clear that our simple model only yields rough estimates and its accuracy should not be overrated. A refined model, which would yield more precise estimates should take into account Fourier frequency dependent correlations, locally dependent environmental parameters, and dynamical effects such as the water vapor diffusion into the fiber cladding. This would go beyond the scope of this paper. However, the model suffices to provide better insights into the relative orders of magnitude of the various effects and also demonstrates the stability enhancement gained by our end-to-end approach.

In the CP case, two fields at different wavelengths propagate on a common path and only a differential contribution remains. Thus in this case, we also assume full correlations between the individual geometric length fluctuations, but take into account that the refractive indices are wavelength-dependent due to chromatic dispersion. Hence the overall noise contribution for two fields at $\lambda_1$ and $\lambda_2$ propagating on a common path of length $L$ is
\begin{equation}
y_{\zeta}(\lambda_1)- y_{\zeta}(\lambda_2) = \frac{1}{c}
\left[n(\lambda_2)-n(\lambda_1)\right] \frac{\partial L}{\partial\zeta} 
\frac{\mathrm{d}\zeta}{\mathrm{d}t} +
\frac{L}{c}
\left[\frac{\partial n(\lambda_2)}{\partial\zeta} - \frac{\partial n(\lambda_1)}{\partial\zeta} \right]
\frac{\mathrm{d}\zeta}{\mathrm{d}t}.
\label{eq:cp_contrib}
\end{equation}
where the second term is a term due to dispersion fluctuations induced by parameter $\zeta$. In a simple model we assume that 
\begin{equation}
\frac{\partial n(\lambda_2)}{\partial\zeta}=
\frac{n(\lambda_2)-1}{n(\lambda_1)-1}  \,\frac{\partial n(\lambda_1)}{\partial\zeta},
\label{eq:prop_disp}
\end{equation}
which means that the spectral dependence of $\partial n / \partial \zeta$ is proportional to the deviation of the refractive index from its value 1 in vacuum, where the effect is zero.
Plugging Eq.~(\ref{eq:prop_disp}) into Eq.~(\ref{eq:cp_contrib}) yields:
\begin{equation}
y_{\zeta}(\lambda_1)- y_{\zeta}(\lambda_2) = \frac{1}{c}
\left[n(\lambda_2)-n(\lambda_1)\right] \frac{\partial L}{\partial\zeta} 
\frac{\mathrm{d}\zeta}{\mathrm{d}t} +
\frac{L}{c}\frac{n(\lambda_2)-n(\lambda_1)}{n(\lambda_1)-1}\,
\frac{\partial n}{\partial\zeta}\,\frac{\mathrm{d}\zeta}{\mathrm{d}t}.
\label{eq:disp}
\end{equation}

We measured typical temperature, pressure and relative humidity time series in our lab and model the frequency transfer setup with path types and lengths approximating those in the NCP and CP experimental setups. 

Besides the contributions from environmental parameter fluctuations, the contribution from the beat detection is also considered. The beat signals consist of an RF carrier on a white noise floor with a phase noise density $S_\varphi\approx-80\,\mathrm{dBrad^2/Hz}$. This corresponds to a modified Allan deviation of
\begin{equation}
\mathrm{mod}\,\sigma_y(\tau)= \frac{\sqrt{3/2S_\varphi}}{2\pi \nu_0}\tau^{-3/2}\approx10^{-19}(\tau / \mathrm{s})^{-3/2}.
\label{eq:modADEV}
\end{equation}
However, the counters used for counting the beat frequencies (K\&K FXE80) yield a significantly higher modified Allan deviation due to an aliasing effect: These counters are essentially phase recorders~\cite{kra04a}, which sample the phase at a sampling rate of $f_\mathrm{s}=1$~kHz. If present, phase noise at higher Fourier frequencies is aliased to the 1 kHz bandwidth~\cite{ver98}. In our experiments, the signals analyzed with the counters show white phase noise with density $S_\varphi$ up to a cutoff frequency of $f_\mathrm{h}\approx 1$~MHz given by the tracking filter. Hence, the $\Lambda$-weighted frequency average using these data leads to a modified Allan deviation that is increased by a factor 
\begin{equation}
k= \sqrt{\frac{2f_h}{f_\mathrm{s}}}
\label{eq:aliasing}
\end{equation}
compared to the true modified Allan deviation~\cite{ver98a}. Furthermore, multiple counted beat signals enter into the transfer process (two for the CEO-free comb, three for the reference comb), which leads to an additional factor given by the square root of the number of involved beat signals if we assume that they contribute equally.

We have also measured separately the additional noise introduced by the analog tracking oscillators. For this purpose, we have replaced the beat signals by synthesizer-generated signals with a $S_\varphi\approx-80\,\mathrm{dBrad^2/Hz}$ white phase noise floor, which was added using cascaded noisy RF-amplifiers. Two such signals with independent noise floors were sent through two tracking oscillators with tracking bandwidths of $f_h \approx 1$~MHz. Their output signals were counted by the K\&K counters. Subsequently, the modified Allan deviation estimate was computed from the difference of the $\Lambda$ counted frequency values, yielding an estimate for the noise due to two tracking oscillators, which can be rescaled taking into account the actual number of tracking oscillators involved in an experiment.

\begin{figure}[htbp]
\centering
\includegraphics[width=\linewidth]{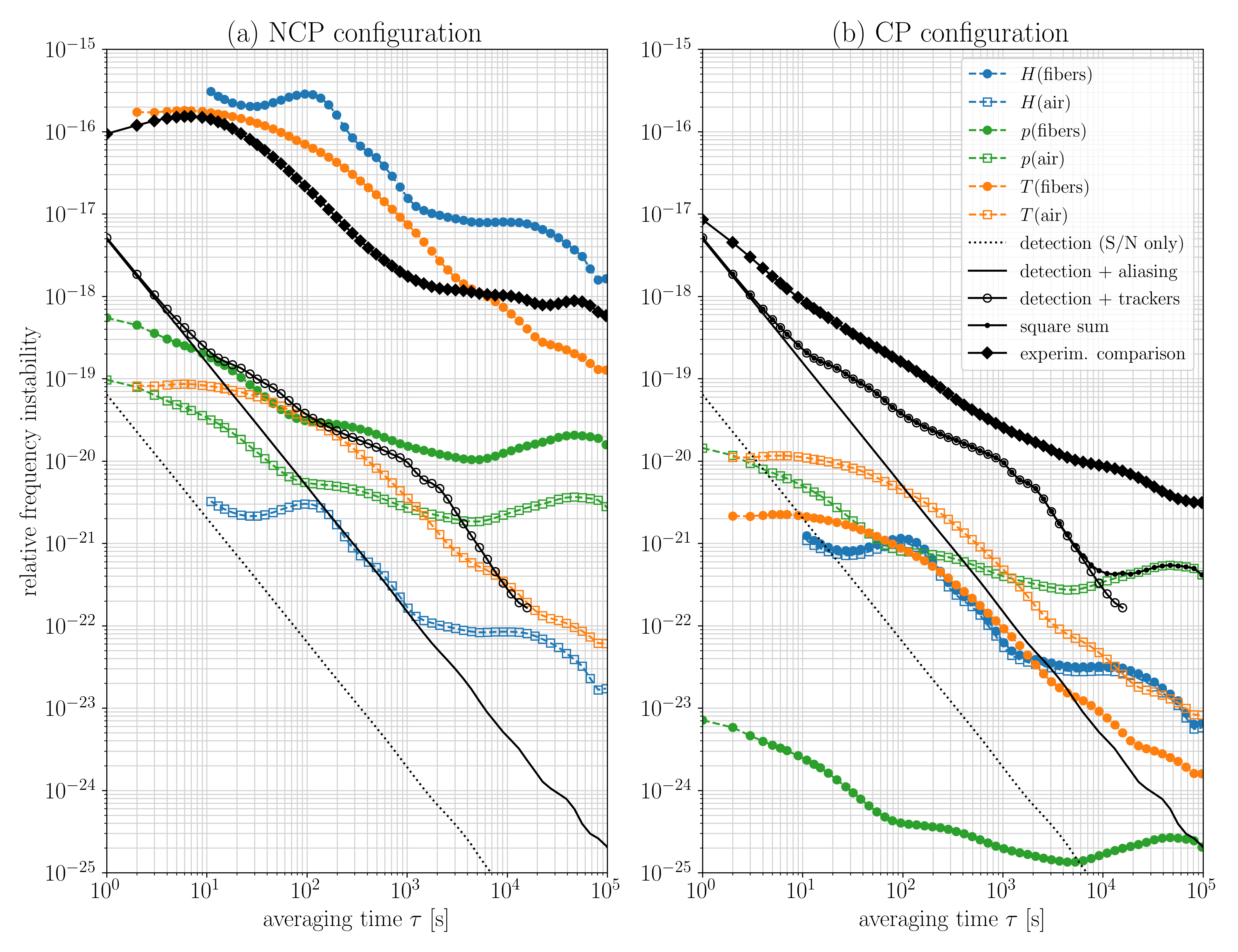}
\caption{Estimated contributions from various noise processes. Open square symbols: Contributions from free space air paths; Filled circle symbols: Fiber paths. Blue: Humidity, green: Pressure, orange: Temperature. (a) NCP case, i.e. including  non-compensated paths. (b) CP case, i.e. end-to-end path length compensated. See text for details.}
\label{fig:estim}
\end{figure}
The results are shown in Fig.~\ref{fig:estim}(a) for the NCP case and (b) for the CP case, with the lengths reflecting the situation in our experimental setup, as given in table~\ref{table2}. In both cases, we plot the contributions to the relative frequency instability from the various environmental parameter variations. Furthermore, we show both the true modified Allan deviation corresponding to the detection noise level of $S_\varphi\approx -80\,\mathrm{dBrad^2/Hz}$ (Eq.~(\ref{eq:modADEV})) and the instability taking into account aliasing in the frequency counters (Eq.~(\ref{eq:aliasing})), which is a more realistic estimate for comparison with our experimental results. In the CP case, we also plot the square sum of all contributions discussed (black circles). Furthermore, we plot again the results of the combined excess noise measurements taken from Fig.~\ref{fig:results}, indicated by black diamond symbols.

With the NCP setup, fiber dilations through relative humidity variations appear as the largest contribution in our estimation. However, as mentioned before, the estimations for humidity variations have to be interpreted with care and are very likely overestimated. This particularly applies to short averaging times below 1000~s, as the water vapour indiffusion time into the coating is on the order of an hour or even more for fibers with polymer jacket.
The other main contribution in the NCP case is due to temperature fluctuations mediated through the temperature dependence of the refractive index in the fibers, followed by pressure- and temperature-induced refractive index fluctuations in the free air sections.

The measured NCP frequency transfer instability resembles the estimated contributions from temperature and relative humidity variations in Fig.~\ref{fig:estim}(a), although it is clear that they should only be understood as order-of-magnitude estimates.

All path-length-induced noise contributions in the CP setup are estimated to be common-mode rejected to instability levels around $10^{-20}$~@~1~s and below $10^{-21}$ @ $10^5$~s (Fig~\ref{fig:estim}(b)).  At intermediate averaging times (30~s~$< \tau <$~1000~s) temperature-induced air refractive index fluctuations in the free space sections (see table~\ref{table2} for the lengths) are estimated to be the major path-length induced contributions in the CP case. 
At longer averaging times, the noise is dominated by air pressure fluctuations in the free space sections in the low $10^{-21}$ range at $\tau>1000$~s. The temperature and pressure contributions in the free space sections could be further reduced by evacuating the air from the critical paths as in~\cite{nic14}.

At small averaging times $\tau<10$~s, the aliased detection noise dominates the estimated instability in the CP case. Since this contribution is only due to the data acquisition equipment used, it is not relevant for an actual frequency transfer. Anyhow, it could be reduced by the factor in Eq.~(\ref{eq:aliasing}) to its real value of less than $10^{-19} @ 1$~s by using a counter which allows suitable anti-aliasing filters. Further approaches would be to reduce the detection noise by gating the cw fields generating the beat notes, such that noise during the time between the comb pulses is suppressed \cite{des13}, or to suppress amplitude noise by balanced detection \cite{rue11}.

At averaging times 10~s~$< \tau<$~5000~s, the noise from the tracking oscillators dominates. The noise in the tracking oscillators is ascribed to a temperature dependence of their voltage-controlled oscillators combined with finite gain of the tracking phase-locked loop. This can be reduced by total elimination of the tracking oscillators, or by digitally implementing their function in a field-programmable gated array (FPGA)~\cite{tou18}. 

Besides a detail near $\tau=10000$~s, the CP frequency transfer instability  is qualitatively correctly predicted by the estimate, while quantitatively it is larger by a factor of roughly 2. Keeping in mind that the estimates have a rather large uncertainty of up to an order of magnitude, the agreement between the estimate and the observed CP instability is acceptable.

Further processes, which were not taken into account when determining the estimates, but could explain the small observed discrepancy are: 

(i) The fact that the long data sets had several gaps which break the phase-coherence and hence also influence the instability derived from the data set.

(ii) Noise during the spectral broadening in the nonlinear fibers \cite{cor03, hav04a,lie19} and during frequency doubling~\cite{her19}. Further investigations will be necessary to single out these contributions. 

\begin{table}[htbp]
\centering
\caption{\bf Order-of-magnitude estimates for the various fractional frequency ratio offsets $y_\mathrm{offs}$ due to linear drifts between DUT and reference system during a measurement time of 6~days. As in the case of the instabilities, the offset resulting from humidity in fibers is probably over-estimated using the humidity coefficient in table~\ref{table1}. For this reason, this coefficient has been reduced by a correction factor of 10.}
\begin{tabular}{ |c|c|c|c| }
\hline 
contribution from & path type & NCP & CP \\ 
\hline
\multirow{2}{*}{$\mathrm{d}T/\mathrm{d}t= 5\times10^{-7}\,\mathrm{K/s}$} & air & $5\times10^{-23}$ & $7\times10^{-24}$\\
                     & fiber & $1\times10^{-19}$ & $1\times10^{-24}$\\
\multirow{2}{*}{$\mathrm{d}p/\mathrm{d}t= 5\times10^{-3}\,\mathrm{Pa/s}$} & air & $2\times10^{-21}$ & $4\times10^{-22}$\\
                     & fiber & $1\times10^{-20}$ & $2\times10^{-25}$\\
\multirow{2}{*}{$\mathrm{d}H/\mathrm{d}t= 4\times10^{-6}\, \%/\mathrm{s}$} & air & $2\times10^{-23}$ & $7\times10^{-24}$\\
                     & fibers & $2\times10^{-19}$ & $9\times10^{-25}$\\
tracking filters & & $3\times10^{-23}$ & $3\times10^{-23}$\\
AM-PM in PDs & & $2\times10^{-25}$ & $2\times10^{-25}$\\
\hline
square sum & & $2\times10^{-19}$ & $4\times10^{-22}$\\
\hline
\end{tabular}
\label{table3}
\end{table}
Besides by the instabilities, the frequency transfer performance is characterized by the offset between the frequency ratios measured at the DUT and reference system. In the ideal case, this frequency offset is zero, i.e. the measurement should yield the same frequency ratio at the DUT and reference comb. A real measurement has a finite measurement time and according to equation~(\ref{eq:total_diff}), a difference of the product between the sensitivity and the linear drift during the measurement time of the parameters $\zeta(t)$ in the DUT and reference system thus leads to a nonzero offset. The offsets estimated by using the coefficients in table~\ref{table1} and lengths in table~\ref{table2} for the effects in fibers, the Edl\'en formula in air, and linear drifts determined from environmental parameter time series are shown in table~\ref{table3}. For the contribution of the phase tracking filters we assume that it is mainly caused by a temperature drift. Under this assumption, we can estimate the offset contribution of the trackers by scaling of the offset due to temperature drift with the ratio between the instabilities arising from the trackers and from the temperature-induced path length variations. The contribution from photodetection arises due to AM-PM conversion. For a worst case estimate, we assume a differential linear power drift of $10^{-4}$ during the measurement time of roughly 6~days, and a typical AM-PM coefficient of 1~rad$/(\mathrm{d}P/P)$~\cite{tay11}.

\section{Conclusion and outlook}
We have introduced a technically simple scheme for the generation of beat signals between cw light fields and an optical frequency comb, which allows end-to-end suppression of optical path length fluctuation induced instabilities during frequency transfer via optical frequency combs. Comparing the frequency transfers based on this scheme at a CEO-free comb and at a second, independent comb with $f-2f$ CEO frequency detection, we demonstrated a factor of 200 suppression of excess noise with respect to a setup employing two separate branches at one of the two combs. A residual transfer combined instability of the ref and DUT system of less than $8 \times 10^{-18} @ 1$~s and $3 \times 10^{-21} @ 10^5$~s were observed along with a residual zero-compatible offset of $9.4 \times 10^{-22}$. This experiment is also the first demonstration that difference-frequency-generation (DFG) based Er-fiber comb generators are suitable for optical frequency transfers at such a level. Furthermore, we presented a simple model for the estimation of individual noise contributions and found that temperature- and relative humidity fluctuations dominate the optical path length fluctuations. Furthermore, we identified two limitations due to the involved electronics: First, aliasing at the counters leads to a factor of $\approx 100$ elevation of the influence of white detection phase noise in the frequency instability (modified Allan deviation), and second we found that the employed analog tracking oscillators significantly contribute to the excess noise. We made suggestions how these technical limits could be overcome by implementing the RF signal processing in FPGA technology, such that the frequency transfer via a frequency comb will be better than the instability even of of the best reference lasers expected in the next few years.

\section*{Funding}
Part of this work has received funding in the 15SIB03 OC18 and 18SIB05 ROCIT projects from the EMPIR programme co-financed by the Participating States and from the European Union's Horizon 2020 research and innovation programme.

\section*{Disclosures}
The use of trade names in this article is necessary for completeness and does not constitute an endorsement by Physikalisch-Technische Bundesanstalt.

\noindent RW: TOPTICA Photonics (E), TP, FR: TOPTICA Photonics (E, I)

\section*{Acknowledgment}
The authors thank T. Legero and S. H\"afner for providing the cavity-stabilized source and target light fields. 

\bibliography{TOP_RPLC}

\end{document}